\documentclass[twocolumn,twoside]{IEEEtran2e}
\usepackage{latexsym,amssymb,cite}

\newcommand{\onemat}[0]{{\mathbf 1}}
\newcommand{\SL}[0]{{\rm SL}}
\newcommand{\GL}[0]{{\rm GL}}
\newcommand{\DFT}[0]{{\rm F}}
\newcommand{\C}[0]{{\mathbb{C}}}
\newcommand{\F}[0]{{\mathbb{F}}}

\newcommand{\Q}[0]{{\mathbb{Q}}}
\newcommand{\R}[0]{{\mathbb{R}}}
\newcommand{\Z}[0]{{\mathbb{Z}}}

\newcommand{\mat}[4]{\left(\begin{array}{rr}#1&#2\\#3&#4\end{array}\right)}

\newcommand{\nix}[1]{}

\def\ket#1{\left|#1\right>}

\def\trace{\mathop{{\rm tr}}\nolimits}
\def\tr{\mathop{{\rm tr}}\nolimits}

\newtheorem{theorem}{Theorem}

\newtheorem{lemma}[theorem]{Lemma}

\newtheorem{example}[theorem]{Example}

\begin{document}
\title{On the Monomiality of Nice Error Bases}

\author{Andreas Klappenecker\thanks{The work by A.K. was supported 
in part by NSF grant EIA 0218582 and a Texas A\&M TITF grant. The 
work by M.R. was supported in part by EC grant IST-1999-10596 (Q-ACTA), 
by CSE and MITACS.}
\thanks{A. Klappenecker is with the Department of Computer Science, 
Texas A\&M University, College Station, TX 77843-3112, USA
(e-mail: klappi@cs.tamu.edu)}
and Martin R\"otteler\thanks{M. R\"otteler is with the
Department of Combinatorics and Optimization, University of Waterloo, Waterloo, Ontario, Canada, N2l 3G1 (e-mail: roettele@iqc.ca)}
}

\maketitle

\begin{abstract}
Unitary error bases generalize the Pauli matrices to higher
dimensional systems. Two basic constructions of unitary error bases
are known: An algebraic construction by Knill, which yields nice error
bases, and a combinatorial construction by Werner, which yields
shift-and-multiply bases.  An open problem posed by Schlingemann and
Werner relates these two constructions and asks whether each nice
error basis is equivalent to a shift-and-multiply basis. We solve this
problem and show that the answer is negative. However, we also show
that it is always possible to find a fairly sparse representation of a
nice error basis.
\end{abstract}

\begin{keywords}
Pauli matrices, unitary error bases, monomial representations, 
Hadamard matrices, Latin squares.
\end{keywords}

%
%

\section{Introduction}

Unitary error bases are important primitives in quantum information
theory.  They form the basis of quantum error-correcting codes,
teleportation, and dense coding schemes.  A unitary error basis is by
definition an orthonormal basis of the vector space of complex
$d\times d$ matrices with respect to the inner product
$\langle A, B \rangle = 1/d \, {\rm tr}(A^\dagger B)$.  Such bases
have been studied in numerous works, see for instance
\cite{Frucht:31,KR:2002a,Knill:96,Schwinger:60,schwinger01,PZ:88}.
However, surprisingly little is known about their general structure.

Currently, two fundamentally different constructions of unitary error
bases are known: An algebraic construction due to
Knill~\cite{Knill:96}, which yields nice error bases, and a
combinatorial construction due to Werner~\cite{Werner:2001}, which
yields shift-and-multiply bases.
The nice error bases of small dimension have been completely classified
in~\cite{KR:2002a}. A quick inspection of this catalogue shows that
each nice error basis in dimension $d\le 5$ is in fact equivalent to a
basis of shift-and-multiply type.  This motivated Schlingemann and
Werner to formulate the following problem~\cite{schlingemann:2002,werner00}:

{\smallskip\par\em\noindent\leftskip=20pt\rightskip=20pt%
Is every nice error basis equivalent to a basis
of shift-and-multiply type?\par\smallskip}

We will give a precise explanation of the technical terms in the next
section. An affirmative answer to this problem would imply that a nice
error basis can be represented by monomial matrices, which have only
one nonzero entry in each row and in each column. 

Do nice error bases really have such a simple structure?  The answer
is no, as we will show in this correspondence.  However, we will prove
that each nice error basis is equivalent to a unitary error basis
where at least half of the entries in the basis matrices are zero.  In
that sense, the nice error bases are simpler than one would expect.

We will recall the definition and some properties of nice error bases
and shift-and-multiply bases in the next section.  We introduce a
notion of equivalence for unitary error bases in Section~III. We show
that there exist shift-and-multiply bases which are not nice error
bases. We then go on to prove that there exists an infinite number of
nice error bases, which are not of shift-and-multiply type. We
construct an explicit counterexample in dimension 165 in Section~V.

\textit{Notations.}  We denote by $\C$ the field of complex numbers,
by $\Z$ the ring of integers, by $\Z_d$ the ring of integers
modulo~$d$. The group of unitary $d\times d$ matrices is denoted by
${\mathcal{U}}(d)$, the general linear group by $\GL(d,\C)$.

\nix{The nice error bases proved to be
very useful in the construction of quantum error correcting
codes~\cite{CRSS:98,Gottesman:96,KR:2002b,KR:2002c}.

This basis is both a nice error basis and a shift-and-multiply basis. 
It is known that in two dimensions the basis (\ref{pauli}) is unique
up to equivalence \cite{VW:2000}.
}

%
%

\section{Construction of Unitary Error Bases}
A unitary error basis is a set $\mathcal{E}$ of $d^{\,2}$ unitary $d\times
d$ matrices such that $\tr(E^\dagger F)=0$ for all distinct $E,F\in
\mathcal{E}$. The set ${\mathcal{P}}$ of Pauli matrices provides the most well-known example: 
$$ {\mathcal{P}} = \left\{
\pmatrix{ 1 & 0 \cr 0 & 1},
\pmatrix{ 0 & 1 \cr 1 & 0},
\pmatrix{ 1 & \phantom{-}0 \cr 0 & -1},
\pmatrix{ 0 & -i \cr i & \phantom{-}0}
\right\}.
$$
Unitary error bases generalize this example to arbitrary
dimensions. The nonbinary case is more interesting, since there exist
different, non-equivalent, error bases. We review in this section the
constructions of unitary error bases by Knill and by Werner.

\subsection{Equivalence}
Let ${\cal E}$ and ${\cal E'}$ be two unitary error bases in $d$ dimensions. 
We
say that ${\cal E}$ and ${\cal E'}$ are equivalent, in signs ${\cal E}
\equiv {\cal E'}$, if and only if there exist unitary matrices 
$A, B \in {\cal U}(d)$ and constants $c_E\in {\mathcal{U}}(1)$, $E\in \mathcal{E}$, 
such that 
$$ {\mathcal{E}}' = \{c_E A E B : E \in {\cal E}\}.$$
One readily checks that $\equiv$ is an equivalence relation. 

\begin{lemma}
Any unitary error basis in dimension $2$ is equivalent to the Pauli basis. 
\end{lemma}
\begin{proof} Let 
${\mathcal{A}}=\{ A_1,A_2,A_3,A_4\}$ be an arbitrary unitary error
basis in dimension 2.  This basis is equivalent to a basis of the form
$\{ \onemat_2, \mbox{diag}(1,-1),B_3,B_4\}$. The diagonal elements of
$B_3$ and $B_4$ are necessarily zero, because of the trace
orthogonality relations.  We may assume that $B_3$ and $B_4$ are of
the form $B_3=\mbox{antidiag}(1,a)$ and $B_4=\mbox{antidiag}(1,-a)$,
where $a=\exp(i\phi)$ for some $\phi\in \R$, since we are allowed to
multiply the matrices with scalars.  Conjugating the basis elements
with the matrix $\mbox{diag}(1,\exp(-i\phi/2))$ yields the matrices of
Pauli basis up to scalar multiples, hence ${\mathcal{A}}\equiv
{\mathcal{P}}$.
\end{proof}

\subsection{Nice error bases}
Let ${G}$ be a group of order $d^2$ with identity element~1.  A {nice
error basis} in $d$ dimensions is given by 
a set ${\cal E}=\{ \rho(g)\in {\cal
U}(n) \,|\, g\in {G}\}$ of unitary matrices such that
\begin{tabbing}
i)\= (iiiii) \= \kill
\>(i)   \> $\rho(1)$ is the identity matrix,\\[1ex]
\>(ii)  \> $\trace\rho(g)=0$ for all $g\in {G}\setminus \{1\}$,\\[1ex]
\>(iii) \> $\rho(g)\rho(h)=\omega(g,h)\,\rho(gh)$ for all $g,h\in
{G}$,
\end{tabbing}
where $\omega(g,h)$ is a phase factor. Conditions (i) and (iii) state
that $\rho$ is a projective representation of the group $G$.

\goodbreak
\begin{lemma}[Knill] 
A nice error basis is a unitary error basis.
\end{lemma}
\begin{proof}
Let ${\cal E}=\{ \rho(g)\in {\cal U}(n) \,|\, g\in {G}\}$ be a
nice error basis. Notice that $\rho(g)^\dagger =
\omega(g^{-1},g)^{-1}\rho(g^{-1})$. Assume that $g,h$ 
are distinct elements of $G$, then $g^{-1}h\neq 1$, 
hence $\tr(\rho(g)^\dagger
\rho(h))=\omega(g^{-1},g)^{-1}\omega(g^{-1},h)\tr(\rho(g^{-1}h))=0$ 
by property (ii) of a nice error basis.
\end{proof}

The next example shows that nice error bases exist in arbitrary dimensions:
\begin{example}\label{cyclicshift}
Let $d\ge 2$ be a integer, $\omega = \exp(2\pi i/d)$.  Let $X_d$
denote the cyclic shift $X_d\ket{x}=\ket{ x-1\bmod d}$, and let $Z_d$
denote the diagonal matrix ${\rm diag}(1, \omega, \omega^2, \ldots,
\omega^{d-1})$. Then ${\cal E}_d := \{X^i_d Z^j_d | (i,j)\in
\Z_d\times \Z_d\}$ is a nice error basis. This has been shown 
by an explicit calculation in \cite{AK:2001}.
\end{example}
\smallskip

\subsection{Shift-and-Multiply Bases}
Recall that a Latin square of order $d$ is a $d\times d$ matrix such
that each element of the set $\Z_d$ is contained exactly once in each
row and in each column. A complex Hadamard matrix $H$ of order $d$ is
a matrix in $\GL(d,\C)$ such that $H_{ik}\in {\mathcal{U}}(1)$, $0\le
i,k<d$, and $H^\dagger H=d\onemat$.

Let ${\mathbf{H}}=(H^{(i)} : 0\le i<d)$ be a sequence of complex
Hadamard matrices, and let $L$ be a Latin square $L$ of order~$d$.
A shift-and-multiply basis ${\mathcal{E}}$ associated with 
$L, \mathbf{H}$ is given by the unitary matrices 
\begin{equation}\label{eq:sm}
E_{ij} = P_j \,{\rm diag}(H^{(j)}_{ik} : 0\le k<d), \quad i,j\in \Z_d, 
\end{equation}
where $P_j$ denotes the permutation matrix with entries defined by
$P_j(L(j,k),k)=1$, for $0\le k<d$, and $0$ otherwise.  In short,
$E_{ij}$ is determined by the $i$th row of the $j$th Hadamard matrix
$H^{(j)}$, and by the entries of the $j$th row of the Latin square
$L$; briefly $E_{ij}\ket{k}=H_{ik}^{(j)}\ket{L(j,k)}$.

If all matrices in $\mathbf{H}$ are equal to a single
Hadamard matrix $H$, then we refer to this basis as the
shift-and-multiply basis associated with $L,H$.

\begin{lemma}[Werner]
A shift-and-multiply basis is a unitary error basis. 
\end{lemma}
\begin{proof} We have to show that 
${\rm tr}(E_{ij}^\dagger E_{kl}) = 0$ when $(i,j)\neq (k,l)$.  If
$j\neq l$, then the matrix $P_j^\dagger P_l^{\phantom{\dagger}}$ has a
vanishing diagonal, whence ${\tr}(E_{ij}^\dagger E_{kl}) = 0$ for any
choice of $i$ and $k$. If $j=l$ and $i\neq k$, then ${\rm
tr}(E_{ij}^\dagger E_{kj})$ is equal to the inner product of the $i$th
and $k$th row of the complex Hadamard matrix $H^{(j)}$, hence ${\rm
tr}(E_{ij}^\dagger E_{kj})=0$. 
\end{proof}
\smallskip

\begin{example}
The nice error basis ${\cal E}_d$ in Example~\ref{cyclicshift} is a
shift-and-multiply basis. Indeed, choose the Latin square 
$L=(j-i\bmod d)_{i,j\in \Z_d}$ and the complex Hadamard matrix 
$H=(\omega^{k\ell})_{k,\ell\in \Z_d}$, with $\omega=\exp(2\pi i/d)$.
For example, if $d=3$, then 
\[
L := 
\left(
\begin{array}{rrr}
0 & 1 & 2  \\
2 & 0 & 1 \\
1 & 2 & 0  
\end{array}
\right),\;
H =
\left(
\begin{array}{lll}
1 & 1 & 1 \\
1 & \omega & \omega^2 \\
1 & \omega^2 & \omega 
\end{array}
\right),\] where $\omega=\exp(2\pi i/3)$. 
According to equation~(\ref{eq:sm}), 
the basis
matrices $E_{01}$ and $E_{12}$ are respectively given by 
$$ 
E_{01} = \left(\begin{array}{lll}
0 & 1 & 0\\
0 & 0 & 1 \\
1 & 0 & 0
\end{array}\right),\;
E_{12} = \left(\begin{array}{lll}
0 & 0 & \omega^2\\
1 & 0 & 0 \\
0 & \omega & 0
\end{array}\right).
$$
The entries of the middle row of the Latin square $L$ and the first
row of the complex Hadamard matrix $H$ determine the matrix~$E_{01}$. 
\end{example}

\subsection{Abstract Error Groups}

Let ${\cal E}=\{\rho(g): g\in G\}$ be a nice error basis. 
A group~$H$ isomorphic to the group generated by the matrices 
$\rho(g)$ is called an abstract error group of $\cal E$. 

The group $H$ is not necessarily finite.  However, if we multiply the
representing matrices $\rho(g)$ by scalars $c_g$ such that
$c_g\rho(g)$ has determinant~1, then the resulting nice error basis
${\cal E}'=\{ c_g\rho(g): g\in G\}$ is equivalent to ${\cal E}$, and
its abstract error group $H'$ is finite. 

Thus, if we consider a nice error basis up to equivalence, then we may
assume without loss of generality that the associated abstract error
group is finite. 

\begin{example}\label{ex:heisenberg} 
The abstract error group $H_d$ associated with the nice error basis
${\cal E}_d$ from Example~\ref{cyclicshift} is by definition
isomorphic to the group generated by $X_d$ and~$Z_d$. 
An element of the group $\langle X_d, Z_d\rangle$ is of the form 
$\omega^z\!Z_d^y\!X_d^{\vphantom{y}x}$, because $X_dZ_d=\omega Z_dX_d$.
\nix{Note that $(\omega^z\!Z_d^y\!X_d^{\vphantom{y}x})
(\omega^{z'}Z_d^{y'}X_d^{\vphantom{y'}x'})=
\omega^{z+z'+xy'}\!Z_d^{y+y'}\!X_d^{\vphantom{y}x+x'}$.}
Notice that $H_d$ is isomorphic to the unitriangular subgroup of 
$\GL(3,\Z_d)$ given by 
$$ H_d\cong
 \left\{ \left( 
\begin{array}{ccc}
1 & x & z \\
0 & 1 & y \\
0 & 0 & 1
\end{array}
\right) : x, y, z \in \Z_d \right\}.
$$
We prefer to describe the group $H_d$ abstractly by the set of
elements $(x,y,z)\in \Z_d^3$ with composition given by $(x,y,z)\circ
(x',y',z') = (x+x',y+y, z+z'+xy')$, where all operation are modulo
$d$. 
\end{example}
\smallskip

Recall that a finite group~$H$ which has an irreducible representation
of large degree $d=\sqrt{(H\!:\!Z(H))}$ is called a group of central
type.  It has been shown in~\cite{KR:2002a} that a finite group $H$ is
an abstract error group if and only if it is a group of central type
with cyclic center. A somewhat surprising consequence is that an
abstract error group has to be a solvable group.

%
%

\section{Wicked Error Bases}
A unitary error basis, which is not equivalent to a nice error basis,
is called wicked.  We show now that there exist an abundance of wicked
shift-and-multiply bases.

\begin{theorem}\label{smNOTnice}
Let ${\cal E}_\alpha$ be the shift-and-multiply basis associated with $L,H_\alpha$, where  
$$ 
L = 
\left(
\begin{array}{rrrr}
0 & 1 & 2 & 3 \\
3 & 0 & 1 & 2 \\
2 & 3 & 0 & 1 \\
1 & 2 & 3 & 0 
\end{array}
\right)\!,\, 
H_\alpha = 
\left(
\begin{array}{rrrr}
1 & 1 & 1 & 1 \\
1 & 1 & -1 & -1 \\
1 & -1 & e^{i \alpha} & -e^{i \alpha} \\
1 & -1 & -e^{i \alpha} & e^{i \alpha} 
\end{array}
\right). 
$$
If $\alpha \in \Q^\times$, then ${\mathcal{E}}_\alpha$ 
is not equivalent to a nice error basis.
\end{theorem}
\begin{proof}
Suppose there exist $A, B \in {\cal U}(4)$, and scalars
$c_{ij}$ such that the set $\{ c_{ij} A U_{ij} B : i, j = 1,
\ldots, 4 \}$ is a nice error basis. Without loss of generality, we may assume that the group $G$
generated by the matrices $c_{ij} A U_{ij} B$ is finite. 

Notice that  the unitary error basis ${\cal E}_\alpha$ contains the matrices 
$M_1 = {\rm
diag}(1, -1, e^{i \alpha}, -e^{i \alpha})$, and $M_2 = \onemat_4$.
Consequently, 
$(c_1 A M_1 B) (c_2 A M_2 B)^{-1} = c_1
c_2^{-1} A M_1 M_2^\dagger A^\dagger = c_1 c_2^{-1} A M_1 A^{-1}$ is an element of the group~$G$. Since $G$ is finite, it follows that $c_1 c_2^{-1} A M_1 A^{-1}$
and hence also that $c_1 c_2^{-1} M_1$ is of finite order.  Looking at
the individual entries of this matrix this implies that $c_1 c_2^{-1}$ and $c_1 c_2^{-1} e^{i \alpha}$ are roots of unity. 
It follows that $e^{i \alpha}$ would have to be a root 
of unity as well, contradicting the assumption $\alpha \in \Q^\times$.
\end{proof}

\section{Sparsity of Nice Error Bases}
A matrix is said to be monomial if and only if it contains exactly one
nonzero entry in each row and in each column.  If a nice error basis
$\mathcal{E}$ is also a shift-and-multiply basis, then all matrices
$M\in {\cal E}$ are monomial. This does not have to be the case.  However,
our next result shows that a nice error basis is always equivalent to
a fairly sparse error basis:

\begin{theorem}\label{th:notprimitive}
Let $\mathcal{E}$ be a nice error basis.  There exists an equivalent
nice error basis ${\mathcal{E}}_s\equiv {\mathcal{E}}$ such that at
least half of the entries of each matrix $M\in {\mathcal{E}}_s$ are
zero.
\end{theorem}
\begin{proof}
There exists a nice error basis ${\cal E}'\equiv {\cal E}$ such that
the abstract error group $H$ of ${\cal E}'$ is finite.  The group
generated by the matrices of ${\cal E}'$ is an irreducible matrix
group isomorphic to $H$. In other words, $H$ has an irreducible
unitary representation $\rho$ such that 
$$ {\cal E}' = \{ \rho(t)\,|\, t\in T\},$$
where $T$ is a transversal of $H$ modulo $Z(H)$.

Recall that a character $\chi$ of $H$ is said to be induced from an
irreducible character $\psi$ of a proper subgroup $K$ of $H$ if $\chi$
is of the form
$$\chi(x)=\frac{1}{|K|}\sum_{h\in H\atop hxh^{-1}\in K}
\psi(hxh^{-1}).$$ Let $\chi$ be the irreducible character of $H$
corresponding to the representation~$\rho$.  If $\chi$ is induced by
an irreducible character of a proper subgroup of $H$, then 
there exists a base change $A\in {\cal U}(d)$ 
such that at least half of the entries of 
the matrices $A\rho(h)A^{-1}$ are zero.  

Seeking a contradiction, we assume that $\chi$ is not induced by an
irreducible character of some proper subgroup of~$H$. In other words,
we assume that $\chi$ is a primitive character.  Notice that the
degree of $\chi$ is large, $\chi(1)=\sqrt{(H\!:\!Z(H))}$.  It follows
that the multiple $\chi(1)\chi$ of the character $\chi$ is induced
from a character from the center $Z(H)$, see \cite{isaacs76}.  Since
$H$ is an abstract error group, it is in particular a solvable 
group~\cite{KR:2002a}.
It has been shown by Ferguson and Isaacs~\cite{ferguson89} that the
multiple of a primitive character of a solvable group can never be
induced from an irreducible character of a proper subgroup,
contradiction.

It follows that 
${\cal E}_s= \{A\rho(t)A^{-1}\,|\, t\in T\}$ 
is a nice error basis, 
and half of the entries of $A\rho(t)A^\dagger$ are zero. By construction, 
${\cal E}\equiv {\cal E}'\equiv {\cal E}_s$, hence 
${\cal E}\equiv {\cal E}_s$ as claimed.
\end{proof}

%
%

\section{Nonmonomial Abstract Error Groups}\label{naeg}
In this section, we will finally answer the question raised by
Schlingemann and Werner:
\begin{theorem}\label{th:counterex}
There exist nice error bases which are not equivalent to 
bases of shift-and-multiply type. 
\end{theorem}

We will prove this result with the help of abstract error groups.
The following result will play a key role in our proof: 

\begin{theorem}[Dade, Isaacs]\label{th:dade}
There exist a group $H$ of central type with cyclic center, which has
a nonmonomial irreducible character $\chi$ of degree
$\chi(1)=\sqrt{(H\!:\!Z(H))}$.
\end{theorem}

\def\proof{\noindent\hspace{2em}{\it Proof of Theorem~\ref{th:counterex}: }}
\begin{proof} 
We actually show a stronger statement, namely that there
are nice error bases which are not equivalent to monomial bases. 

By Theorem~\ref{th:dade}, there exists an abstract error group $H$
that has a non-monomial irreducible unitary representation $\rho$ of
degree $d=\sqrt{(H\!:\!Z(H))}$.  Denote by ${\cal E}$ a nice error
basis associated with $\rho$, that is,
$$ {\cal E} = \{ \rho(t)\,|\, t\in T\}$$ where $T$ is a transversal of
$H$ modulo $Z(H)$, with $1\in T$.  
Since $\rho$ is nonmonomial, it is impossible to
find a base change $A$ such that $A \rho(t) A^\dagger$ is monomial for
all $t\in T$.  We show next that this property is even preserved with
respect to the equivalence~$\equiv$.

Seeking a contradiction, we suppose that there exist unitary matrices
$A, B$ and scalars $c_t$ such that $c_t A \rho(t) B$ is a monomial
unitary error basis. Since the identity matrix $\onemat_d=\rho(1)$ is
part of the nice error basis, we can conclude that the matrix $C = c_1
A B$ is monomial. But $c_t A \rho(t) B = c_t A \rho(t) (A^\dagger A) B
= c_t/c_1 A \rho(t) A^\dagger C$ shows that the resulting equivalent
error basis is nonmonomial. Indeed, among the matrices $A \rho(t)
A^\dagger$ is at least one nonmonomial matrix $U$. Multiplying $U$
with the monomial matrix $C$ and the scalar prefactor $c_t/c_1$ cannot
result in a monomial matrix, leading to a contradiction.
\end{proof}
\def\proof{\noindent\hspace{2em}{\it Proof: }}
\smallskip

\textit{Remark.} We have shown in the proof of
Theorem~\ref{th:notprimitive} that an irreducible character $\chi$ of
large degree of an abstract error group~$H$ is always induced from an
irreducible character $\psi$ of a proper subgroup. The essence of
Theorem~\ref{th:dade} is that in general we cannot choose~$\psi$ to be
a linear character.

In the next section, we want to construct an explicit example of a
nice error basis that is not equivalent to a shift-and-multiply basis.
We will need an explicit example of a group~$H$ satisfying the
assumptions of Theorem~\ref{th:dade} for that purpose.
Theorem~\ref{th:dade} was independently proved by Everett Dade and by
Martin Isaacs, but unfortunately their results remained unpublished.
Our exposition is a variation of Dade's approach, which we include
here with the kind permission of Professor Dade.

\nix{
We take Heisenberg groups as a starting point to
construct an abstract error group of a nice error basis, which is not
equivalent to a shift-and-multiply basis.  We recall some basic facts
about semi-direct products of groups, the Heisenberg groups, and their
automorphisms. These facts which will allow us to construct this group.
}
\subsection{Semidirect Products} 
Let $N, H$ be finite groups, and let $\varphi$ be a group homomorphism
from $H$ to $\mbox{Aut}(N)$.  Recall that the (outer) semidirect
product $G=N\rtimes_\varphi H$ is a group defined on the set $N\times
H$, with composition given by
$(n_1,h_1)(n_2,h_2)=(n_1\,\varphi(h_1)(n_2),h_1h_2)$.  If the center
of $H$ acts trivially on $N$, then the center of the semidirect
product is given by $Z(G)=Z(N)\times Z(H)$.  A detailed discussion of
semidirect products can be found for instance in~\cite{AB:95}.

\subsection{Automorphisms of the Heisenberg group $H_p$}
Let $p$ be an odd prime. The Heisenberg group $H_p$ defined in
Example~\ref{ex:heisenberg} has $p^3$ elements.  Recall that the
special linear group $\mbox{SL}(2,\F_p)$ is a matrix group of order
$(p+1)p(p-1)$, which is generated by the matrices
$$ \alpha=\mat{0}{-1}{1}{0}\quad\mbox{and}\quad\beta=\mat{1}{0}{1}{1}.$$ 
The special linear group acts as an automorphisms group on the Heisenberg
group $H_p$. Indeed, define 
$$\begin{array}{lcl}\alpha(x,y,z)&=& (-y,x,z-xy),\\[1ex]
\beta(x,y,z) &=& (x,x+y,z+ \frac{(p+1)}{2}\,x^2).
  \end{array}
$$
It is straightforward to check that $\alpha, \beta\in \mbox{Aut}(H_p)$. 
We will construct the abstract error group by semidirect products of
Heisenberg groups. We will need some more detailed knowledge about
these automorphisms to tailor this construction to our needs.

Recall that a matrix~$M$ is said to act irreducibly on a vector space
$V$ if and only if $\{0\}$ and $V$ are the only $M$-invariant
subspaces of $V$.

\begin{lemma}\label{irredAction}
Let $r, p$ be odd prime numbers such that $r|(p+1)$.  The group
$\SL(2,\F_p)$ contains matrices of order~$r$, and all such matrices
act irreducibly on $\F_p\times \F_p$. 
\end{lemma}
\begin{proof} 
It is known that the group $\SL(2,\F_p)$ has a cyclic subgroup of
order $p+1$, hence contains matrices of order~$r$,
see~\cite[p.~42]{Gorenstein:80}.  Let $g\in \SL(2,\F_p)$ be an element
of order $r$. The subgroup $R=\langle\, g\,\rangle$ of order $r$ acts
on the vector space $V=\F_p\times \F_p$. The orbit length $|Rv|\in
\{1,r\}$ for all $v\in V$. If we denote by $U$ the centralizer of $R$,
then $r$ divides $|V|-|U|=p^2-p^k$, where $k\in\{0,1\}$.  Since $r$
does not divide $p-1$ or $p$, it follows that $k$ has to be 0. Thus,
$0\in V$ is the only fixed point. It follows that $\{0\}$ and $V$ are
the only $R$-invariant subspaces of $V$. Indeed, suppose that $V'$ is
an $R$-invariant subspace with $p$ elements, then the number of
elements in $V'-\{0\}$ must be a multiple of $r$, contradicting
$r\nmid (p-1)$.
\end{proof}
\smallskip

\noindent\textit{Remark:} The preceding lemma also follows from
Theorem~3.5 in Hering~\cite{Hering:74}.

\subsection{A nonmonomial nice error group}
We take now the Heisenberg groups as a starting point to construct an
abstract error group $G$ of a nice error basis, which is not
equivalent to a shift-and-multiply basis.

Let $p, q,$ and $r$ be three distinct odd primes such that $r$ divides $p+1$
and $q+1$. Define
$$G=(H_p\times H_q) \rtimes_\varphi H_r,$$ 
where the action $\varphi$ is chosen such that $\varphi(g)$  
acts trivially on $H_p\times H_q$ for all $g\in Z(H_r)$.

\begin{lemma}
The group $G$ is of central type with cyclic center. The center $Z(G)$
is of order $pqr$.
\end{lemma}
\begin{proof} The three Sylow subgroups $H_p$, $H_q$, and $H_r$ of the
group $G$ are of central type. Since the center of $H_r$ acts 
trivially on $H_p\times H_q$, we have $Z(G)=Z(H_p)\times
Z(H_q)\times Z(H_r)$.  In particular, $H_\ell\cap Z(G)=Z(H_\ell)$ for
$\ell\in \{p,q,r\}$.  It follows from Theorem 2 of \cite{DJ:69} that $G$
is a group of central type.

The center of a Heisenberg group $H_d$ is given by the cyclic subgroup
$Z(H_d)=\{(0,0,z)\,|\, z\in \Z_d\}$. The center of $G$ is thus a
direct product of cyclic groups of coprime orders, hence $Z(G)$ is a
cyclic group of order $pqr$.
\end{proof}
\smallskip

We want to choose $\varphi$ such that $G$ does not
contain a subgroup of index $pqr$. If this is the case, then a
character $\chi\in \mbox{Irr}(G)$ of degree
$\chi(1)=\sqrt{(G\!:\!Z(G))}=pqr$ cannot be monomial, because a
monomial character is induced from a linear character of a 
subgroup of index $pqr$. 

\nix{
\noindent\textit{Remark:} We have $| \mbox{Syl}_p(G) |
=1$ and $| \mbox{Syl}_q(G) | =1$. However, $|\mbox{Syl}_r(G)|> 1$.
}

\begin{lemma}\label{conjsubgroup}
If $G$ has a subgroup $H$ of index $(G\!:\!H)=pqr$, then there
exists a conjugate subgroup $K=H^g$ such that
$|K\cap H_p|=p^2$, $|K\cap H_q|=q^2$, and $|K\cap H_r|=r^2$,
\end{lemma}
\begin{proof} The Sylow subgroups $H_p$ and $H_q$ of $G$ are normal, hence
respectively contain the subgroups of order $p^2$ and of order $q^2$
of~$H^g$ for any $g\in G$. The subgroup of order $r^2$ of $H$ is
contained in some Sylow subgroup of $G$.  The claim follows, since the
$r$-Sylow subgroups in $G$ are all conjugate by Sylow's theorem.
\end{proof}

\begin{theorem} 
Let $p,q,r$ be distinct odd primes such that $r$ divides $p+1$ and
$q+1$. It is possible to choose $\varphi$ such that $G=(H_p\times
H_q)\rtimes_\varphi H_r$ is a group of central type that does not
contain a subgroup of index $pqr$.
\end{theorem}
\begin{proof} 
First, we define the action $\varphi$ of $H_r$ on $H_p\times H_q$.
Recall that the Heisenberg group $H_r$ is generated by the two
elements $a=(1,0,0)$ and $b=(0,1,0)$.  Let $A,B\in \mbox{SL}(2,\F_p)$
be matrices of order $r$.  The element $a$ acts with $A$ on $H_p$, and
trivially on $H_q$.  Similarly, the element $b$ acts trivially on
$H_p$, and with $B$ on $H_p$.

Notice that $c=(0,0,1)=aba^{-1}b^{-1}$ generates the center of $H_r$.
It follows that $c$ acts trivially on $H_p\times H_q$, hence $Z(H_r)$
acts trivially on $H_p\times H_q$. An immediate consequence is that
the center of $G$ is given by $Z(G)=Z(H_p)\times Z(H_q)\times Z(H_r)$.

The Sylow subgroups of $G$ are isomorphic to the Heisenberg groups
$H_p$, $H_q$, and $H_r$, whence all Sylow subgroups of $G$ are of
central type. Moreover, the construction of $G$ ensures that the
intersection of a Sylow subgroup $P$ of $G$ with the center $Z(G)$
gives $Z(P)$. It follows from Theorem~2 of \cite{DJ:69} that $G$ is a 
group of central type. 

Seeking a contradiction, we suppose that $G$ has a subgroup of index
$pqr$.  Lemma~\ref{conjsubgroup} shows that $G$ must then
have a subgroup $K$ such that the intersection of $K$ with $H_p$,
$H_q$, and $H_r$ contains $p^2$, $q^2$, and $r^2$ elements,
respectively.

Let $X=\langle a,c\rangle$ and $Y=\langle b,c\rangle$; both are
subgroups of order~$r^2$ of the Heisenberg group $H_r$. The subgroup
$K_r = H_r\cap K$ cannot coincide with both $X$ and $Y$. Suppose that
$K_r\neq X$. Since $H_r=\langle K_r, X\rangle$, the group $K_r$ must
act irreducibly on $H_q/Z(H_q)$. The subgroup $K_q=K\cap H_q$ cannot exist, 
because $K_r$ would have to normalize $K_q/Z(K_q)$, which is impossible.
Similarly, if $K_r \neq Y$, then the subgroup $K_p=K\cap H_p$ cannot exist. 
Therefore, the group $K$ cannot exist. This proves 
that $G$ does not contain a subgroup of index $pqr$. 
\end{proof}

%
%

\section{An Explicit Counterexample}
We take now a closer look at the examples given in the previous
section. Specifically, it is our goal is to make the construction
explicit for the smallest possible choice of parameters. This will
give us a concrete example of a nice error basis in dimension
$165=3\cdot 5 \cdot 11$ that is not equivalent to a shift-and-multiply
basis.

\subsection{A Representation of $H_p$}
Let $p$\/ be an odd prime. The Heisenberg group $H_p$ has an irreducible
representation $\rho_p\colon H_p\rightarrow {\cal U}(p)$, which
associates to an element $(x,y,z)\in H_p$ the matrix
$$ \rho_p((x,y,z)) = \omega^{z}Z_p^y X_p^x.$$ 
Here $\omega$ denotes the primitive root of unity 
$\omega=\exp(2\pi i/p)$, and $X_p$ and $Z_p$ denote the 
generalized Pauli matrices, as defined in Example~\ref{cyclicshift}.

We will derive a faithful irreducible matrix representation of degree
$165$ of the group $G=(H_5\times H_{11})\rtimes_\varphi H_3$ by a
suitable composition of the representations $\rho_3, \rho_5$, and
$\rho_{11}$.

\nix{
\vspace*{2cm}
Here $\omega$ denotes the primitive $d$-th root of unity $e^{2\pi i
/d}$ and $\DFT_d$ denotes the discrete Fourier transform of length $d$
defined by $\DFT_d := \frac{1}{\sqrt{d}} (\omega^{k \cdot l})_{k, l =
0, \ldots, d-1}$.
}

\subsection{Automorphisms of $H_p$}
Recall that a group $G$ is said to be an {\em inner} semidirect
product of the two subgroups $H$ and $N$ if and only if $N$ is a
normal subgroup of $G$ such that $HN =G$ and $H \cap N = \{1_G\}$. 

The matrix group representing $G=(H_5\times H_{11})\rtimes_\varphi
H_3$ is an inner semidirect product. This means that the action of the
automorphism is realized by a conjugation with a matrix. It suffices
to find matrices which realize the action of the generators $\alpha$
and $\beta$ of $\SL(2,\F_p)$. Recall that $\alpha(1,0,0)=(0,1,0)$ and
$\alpha(0,1,0)=(-1,0,0)$. This means we need to find a matrix 
$A\in {\cal U}(p)$ such that 
$$ X_p^A = Z_p\quad\mbox{and}\quad Z_p^A =
X_p^{-1}.$$
 Similarly, the action of the 
automorphism $\beta$ is determined by 
$\beta(1,0,0)=(1,1,(p+1)/2)$ and $\beta(0,1,0)=(0,1,0)$. 
Hence we need to find a matrix $B\in {\cal U}(p)$ such that 
$$ X_p^B = \omega^{(p+1)/2}Z_pX_p\quad\mbox{and}
\quad Z_p^B = Z_p.$$

We can choose $A$ to be the discrete Fourier transform  
$F_p=\frac{1}{\sqrt{p}}(\omega^{k\ell})_{k,\ell=0,\dots, p-1}$, 
with $\omega=\exp(2\pi i/p)$.
Notice that the diagonal matrix 
$D_p = {\rm diag}(\omega^{i (i-1)/2} : 0\le i < p)$ satisfies
$X_p^{D_p} = Z_pX_p$ and $Z_p^{D_p} = Z_p$. It follows that 
the matrix $B$ can be chosen to be $B=D_pZ_p^{(p+1)/2}$.

\subsection{A Nonmonomial Error Basis in Dimension 165}

We need an element of order 3 of $\SL(2,\F_p)$ to specify the action
of $H_3$ on $H_5\times H_{11}$. We can choose for instance the element 
$\gamma = \beta\alpha\beta\alpha$ in $\SL(2,\F_p)$, i.e.,
\[ 
\gamma 
= 
\left( \begin{array}{rr} 1 &  0 \\ 1 & 1 \end{array} \right)
\left( \begin{array}{rr} 0 & -1 \\ 1 & 0 \end{array} \right)
\left( \begin{array}{rr} 1 &  0 \\ 1 & 1 \end{array} \right)
\left( \begin{array}{rr} 0 & -1 \\ 1 & 0 \end{array} \right).
\]
One easily verfies that $\gamma^3=\onemat$. 

The action of $\gamma$ on the matrix representation $\rho_5(H_5)$ of
the Heisenberg group $H_5$ is realized by conjugation with the matrix
$R_5 = D_5Z_5^3 \cdot F_5 \cdot D_5Z_5^3 \cdot F_5\in {\cal U}(5)$.
Similarly, the action of $\gamma$ on the matrix version of $H_{11}$ 
is given by conjugation with $R_{11} = D_{11}Z_{11}^3 \cdot
F_{11} \cdot D_{11}Z_{11}^3 \cdot F_{11}\in {\cal U}(11)$.

Recall that $H_3$ is generated by the two elements $(1,0,0)$ and
$(0,1,0)$. According to the construction of Section~\ref{naeg}, the
action of $H_3$ is chosen such that the generator $(1,0,0)$ acts with
$\gamma$ on $H_5$ and trivially on $H_{11}$, and the generator
$(0,1,0)$ acts trivially on $H_5$ and by $\gamma$ on
$H_{11}$. Explicitly, we obtain the matrix group:
\[ 
\begin{array}{l}
G=\langle
\onemat_3 \otimes X_5 \otimes \onemat_{11},\;
\onemat_3 \otimes Z_5 \otimes \onemat_{11},\;
\onemat_3 \otimes \onemat_5 \otimes X_{11},\\
\phantom{G:=\langle} \onemat_3 \otimes \onemat_5 \otimes Z_{11},\;
X_3 \otimes R_5 \otimes \onemat_{11},\;
Z_3 \otimes \onemat_5 \otimes R_{11}
\rangle.
\end{array}
\]
The group $G$ is an inner semidirect product of the form $H_3 \ltimes
(H_5 \times H_{11})$. Indeed, the subgroup generated by the first four
generators is isomorphic to $N=H_5 \times H_{11}$ since the
irreducible representations of a direct product are given by tensor
products of the irreducible representations of the factors. We have 
$N\lhd G$ because of the choice of  $R_5$ and $R_{11}$. The
complement $H$ of $N$ is given by the group generated by the to
remaining generators. Obviously the intersection $H \cap N$ is trivial
and $G=HN$ by definition.  This shows that $G = H \ltimes N$.

\nix{
This also shows that $|G| = 4492125 = 3^3 5^3 11^3$, that the centre
is of order $Z(G) = 165 = 3\cdot 5 \cdot 11$ and that $G$ itself
defines an irreducible representation.  
}

We obtain a non-monomial nice error basis by choosing a transversal of
$Z(G)$ in $G$. This nice error basis is in particular not equivalent
to a shift-and-multiply basis.

\section{Conclusions}
We have studied unitary error bases for higher-dimensional systems.
We have shown that all such bases are equivalent in dimension 2.  This
changes dramatically in higher dimensions. We have shown that nice
error bases are in general not equivalent to shift-and-multiply bases,
and vice versa. This solves an open problem posed by Schlingemann and
Werner.

\section*{Acknowledgments} 
We thank Everett Dade and Marty Isaacs for
discussions on non-monomial representations of groups of central type. The
construction of the nonmonomial character of extremal degree given in
Section~\ref{naeg} is due to Everett Dade.  We thank Thomas Beth for
his encouragement and support.

\bibliography{monomial}
\bibliographystyle{IEEE}

\end{document}